\documentclass[journal]{IEEEtran}

\usepackage{url}

\usepackage{graphicx}

\usepackage{bbm}

\usepackage{./bs_macros}

\begin{document}

\title{Dynamic Portfolio Cuts: A Spectral Approach to Graph-Theoretic Diversification}

\author{Alvaro Arroyo, Bruno Scalzo, Ljubi$\check{\text{s}}$a Stankovi\'c, \IEEEmembership{Fellow, IEEE}, and Danilo P. Mandic, \IEEEmembership{Fellow, IEEE}\vspace{-3 mm}

\thanks{A. Arroyo is with the Department of Electrical and Electronic Engineering,
	Imperial College London, London SW7 2AZ, U.K. (E-mail: alvaro.arroyo17@imperial.ac.uk).}
\thanks{B. Scalzo is with the Department of Electrical and Electronic Engineering,
	Imperial College London, London SW7 2AZ, U.K. (E-mail: bruno.scalzo-dees12@imperial.ac.uk).}
\thanks{L. Stankovi\'c is with the Faculty of Electrical Engineering, University of Montenegro, Podgorica, 81000, Montenegro. (E-mail: ljubisa@ac.me).}
\thanks{D. P. Mandic is with the Department of Electrical and Electronic Engineering,
	Imperial College London, London SW7 2AZ, U.K. (E-mail: d.mandic@imperial.ac.uk).}}

%\markboth{IEEE SIGNAL PROCESSING LETTERS, VOL (...), NO. (...), 2021}
\maketitle
\vspace{-1mm}
\begin{abstract}
	Stock market returns are typically analyzed using standard regression, yet they reside on irregular domains which is a natural scenario for \textit{graph signal processing}. To this end, we consider a \textit{market graph} as an intuitive way to represent the relationships between financial assets. Traditional methods for estimating asset-return covariance operate under the assumption of statistical time-invariance, and are thus unable to appropriately infer the underlying true structure of the market graph. This work introduces a class of graph spectral estimators which cater for the nonstationarity inherent to asset price movements, and serve as a basis to represent the time-varying interactions between assets through a  \textit{dynamic spectral market graph}.  Such an account of the time-varying nature of the asset-return covariance allows us to introduce the notion of \textit{dynamic spectral portfolio cuts}, whereby the graph is partitioned into time-evolving clusters, allowing for online and robust asset allocation. The advantages of the proposed framework over traditional methods are demonstrated through numerical case studies using real-world price data.
\end{abstract}
\vspace{-1mm}
\begin{IEEEkeywords}
	Augmented complex statistics, financial signal processing, graph cut, nonstationary portfolios, portfolio optimization, graph spectra, vertex clustering
\end{IEEEkeywords}

\IEEEpeerreviewmaketitle
\vspace{-4mm}
\section{Introduction}

\IEEEPARstart{T}{he} asset-return covariance matrix is central to Modern Portfolio Theory (MPT), and underpins the mathematical analysis of financial markets \cite{Markowitz1952}\cite{Akansu2015}\cite{Akansu2016}\cite{Zhang2017}.  Investment strategies typically consider a vector, $\r(t) \in \domR^{N}$, which contains the returns of $N$ assets at a time instant $t$, the $i$-th entry of which is given by \cite{Feng2016}
\vspace{-3mm}
\begin{equation}
r_{i}(t) = \frac{p_{i}(t) - p_{i}(t-1)}{p_{i}(t-1)}
\end{equation}
where $p_{i}(t)$ denotes the value of the $i$-th asset at a time $t$. The mean-variance optimization of portfolios asserts that the optimal weighting vector of assets, $\w \in \domR^{N}$, is obtained as
%\begin{equation}
%\max_{\w} \;\{  \w^{\Trans}\m - \lambda \w^{\Trans}\R\w \} \label{eq:MVO}
%\end{equation}
\begin{equation}
\begin{aligned}[b]
	&\min_{\w} \{\w^{\Trans}\R\w\}\quad \text{s.t.\hspace{2mm}} \w^{\Trans}\m  = \overline{\mu};  \hspace{2mm} \w^{\Trans}\mathbf{1}  = 1
\end{aligned}
\end{equation}
where $\m = \expect{\r} \in \domR^{N}$ is a vector of expected future returns, $\R = \cov{\r} \in \domR^{N \times N}$ is the covariance matrix of returns, $\overline{\mu}$ is the \textit{expected return target}, and the second constraint guarantees full allocation of capital. %, and $\lambda$ is a Lagrange multiplier, also referred to as the \textit{risk aversion} parameter.
Despite strong theoretical foundations behind MPT, one important unresolved issue remains an accurate estimation of matrix $\R$ \cite{Chopra1993}\cite{Rubio2012}\cite{Deshmukh2020}, as well as instability issues associated with its inversion \cite{LopezdePrado2012}\cite{LopezdePrado2016}. Recent work \cite{22} proposes to resolve these issues through the \textit{portfolio cut} paradigm, based on \textit{vertex clustering} \cite{12}\cite{Araghi2019} of the \textit{market graph} \cite{Boginski2003}\cite{Palomar2020}. By segmenting the original market graph into computationally feasible and economically meaningful clusters of assets, schemes such as \textit{hierarchical risk parity} \cite{LopezdePrado2016} or \textit{hierarchical clustering based asset allocation} \cite{Raffinot2017} can be used to effectively allocate capital and generate wealth.

Despite its intuitive nature, the above approaches rest upon an unrealistic assumption of time-invariance of the covariance, $\R$, despite the well established fact that financial markets follow nonstationary dynamics \cite{Cont2001}\cite{Guharay2013}\cite{Tobar2012}.  Furthermore, the use of sample estimators in nonstationary environments has been demonstrated to incur significant information loss, as established by von Neumann's \textit{mean ergodic theorem} \cite{vonNeumann1932} and Koopman's \textit{operator theory} \cite{Koopman1931}. This can be seen by considering an idealised case whereby the asset price returns evolve in time according to $\r(t) = \mathcal{S}\r(t-1)$, with $\mathcal{S}:\domC^{N}\mapsto\domC^{N}$ denoting a unitary \textit{shift operator} in a Hilbert space. The mean ergodic theorem then asserts that the sample mean approaches the orthogonal subspace of $\r(t)$, that is
\vspace{-1mm}
\begin{equation}
\lim_{T \to \infty}\frac{1}{T}\sum_{t=0}^{T-1} \r(t) = \lim_{T \to \infty}\frac{1}{T}\sum_{t=0}^{T-1}\mathcal{S}^{t}(\r(0)) = \lim_{T \to \infty}\frac{1}{T}\sum_{t=0}^{T-1}\mathcal{P}\r(0)
\label{eq:mean_ergodic_theorem}
\end{equation}
where $\mathcal{P}$ is the orthogonal projection onto the null space of $(\mathbf{I}-\mathcal{S})$, for which $\|\mathcal{P}\r(t)\|_{2} \leq \|\r(t)\|_{2}$ holds owing to the Cauchy-Schwarz inequality.

In the context of graph data analytics, the need to account for the evolution of the underlying system dynamics has driven the development of dynamic learning systems, such as \textit{temporal graph networks}  \cite{Xu2020}\cite{Rossi2020}. We proceed a step further, and employ a recently proposed class of spectral estimators for nonstationary signals \cite{Scalzo2021}, to retrieve a time-varying covariance, $\R(t)$, which caters for cyclostationary properties in market data. This serves as a basis to reformulate the definition of graph connectivity matrices of the market graph, in order to allow them to vary with time and account for long-term economic cycles present in the data. Such nonstationary graph signal processing \cite{Ortega2018}  operators allow us to introduce the concept of \textit{dynamic spectral vertex clustering} which serves as a basis for the proposed \textit{dynamic spectral portfolio cut}. We demonstrate that this makes it possible to account for the seasonal correlations between vertices in the market graph, an important feature in the diversification of investment strategies, which is completely overlooked when using existing static graph topologies.

\section{Preliminaries}

\subsection{A Class of Nonstationary Signal Operators}

Consider a time-frequency expansion  \cite{Loeve1977}\cite{Schreier2003} of the asset returns, $\r(t) \in \domR^{N}$, given by
\begin{equation}
\label{eq:spectral_process_expansion}
\r(t) = \int_{-\infty}^{\infty} e^{\jmath \omega t} \bbr(t,\omega) \, d \omega
\end{equation}
where $\bbr(t,\omega) \in \domC^{N} $ is the realisation of a random spectral process at an angular frequency, $\omega$, at a time instant, $t$. The ``augmented form'' of this spectral process is then \cite{Mandic2009}

\begin{equation}
\ubbr(t,\omega) = \left[ \begin{array}{c}
\bbr(t,\omega)	 \\
\bbr^{\ast}(t,\omega)
\end{array} \right] \in \domC^{2N}
\end{equation}
The augmented spectral variable at each time instant is assumed to be \textit{multivariate complex Gaussian distributed}, with its pdf is given by \cite{Scalzo2021}
\begin{equation}
p(\ubbr,t,\omega) \! = \! \frac{ \exp \! \left[ \! - \frac{1}{2} \! \left( \ubbr(t,\omega) \! - \! \ubbm(\omega) \right)^{\Her} \! \ubbR^{-1} \!(\omega) \!  \left( \ubbr(t,\omega) \! - \! \ubbm(\omega) \right) \! \right] }{\pi^{N} \det^{\frac{1}{2}} ( \ubbR(\omega) )}
\end{equation}
where the \textit{augmented} spectral mean and covariance are respectively given by
\begin{alignat}{2}
\ubbm(\omega) & = \expect{\ubbx(t,\omega)} && = \left[ \begin{array}{c}
\bbm(\omega)	 \\
\bbm^{\ast}(\omega)
\end{array} \right] \\
\ubbR(\omega) & = \cov{\ubbx(t,\omega)} && = \left[ \begin{array}{cc}
\bbR(\omega) & \bbP(\omega) \\
\bbP^{\ast}(\omega) & \bbR^{\ast}(\omega)
\end{array} \right]
\end{alignat} 
Owing to the linearity of the Fourier operator, the time-domain counterpart of the spectral variable will also be  multivariate Gaussian distributed, since a linear function of Gaussian random variables is also Gaussian distributed. Hence, the vector of returns, $\r(t)$, is distributed as 
\begin{equation}
\r(t) \sim \Normal{\m(t),\R(t)} \label{eq:pdf_time-varying_Gaussian}
\end{equation}
where $\m(t) \in \domR^{N}$ and $\R(t) \in \domR^{N \times N}$ are respectively the time-varying mean vector and covariance matrix. Of particular interest to this work is the time-varying covariance, defined as \cite{Scalzo2021}
\begin{align}
\R(t) & = \cov{\r(t)} = \expect{\s(t)\s^{\Trans}(t)} \notag\\
& = \int_{-\infty}^{\infty}\!\int_{-\infty}^{\infty} \!\!\! e^{\jmath (\omega-\nu) t} \bbR(\omega,\nu) +  e^{\jmath (\omega+\nu) t} \bbP(\omega,\nu) \; d\omega d\nu
\label{eq:10}
\end{align}
where $\s(t) = \r(t)-\m(t)$ denotes the \textit{centred} returns. Observe that $\R(t)$ represents a sum of \textit{cyclostationary} components, each modulated at an angular frequency, $\omega$. 

\subsection{Compact Spectral Representation}

\vspace{-0.25mm}

In order to discretize the above concept, consider a set of $M$ frequency bins, $\boldomega = [\omega_{1},...,\omega_{M}]^{\Trans}$, which form a discrete frequency spectrum, so that the time-frequency expansion in (\ref{eq:spectral_process_expansion}) therefore becomes
\begin{equation}
\r(t) = \frac{1}{\sqrt{2M}} \sum_{m=1}^{M} \left( e^{\jmath \omega_{m} t}\bbr(t,\omega_{m}) + e^{-\jmath \omega_{m} t}\bbr^{\ast}(t,\omega_{m}) \right) \label{eq:spectral process_expansion_discrete}
\end{equation}
or in a compact form
\begin{equation}
\r(t)  = \uboldPhi(t,\boldomega)\ubbr(t,\boldomega) \label{eq:DFT_compact}
\end{equation}
The term $\uboldPhi(t,\boldomega) \in \domC^{N \times 2MN}$ is referred to as the \textit{augmented spectral basis}, defined as 
\begin{equation}
\uboldPhi(t,\boldomega)  =  \left[ \begin{array}{cc} 
\boldPhi(t,\boldomega) & \boldPhi^{\ast}(t,\boldomega) \end{array} \right]
\end{equation}
\begin{equation}
\boldPhi(t,\boldomega) = \frac{1}{\sqrt{2M}} \left[ 
\begin{array}{ccc}
e^{\jmath \omega_{1} t}\I_{N} & \cdots & e^{\jmath \omega_{M} t}\I_{N} \end{array} \right]
\end{equation}
with $\I_{N} \in \domR^{N\times N}$ as the identity matrix, and $\ubbr(t,\boldomega) \in \domC^{2MN}$ as the \textit{augmented spectrum representation}, given by
\begin{equation}
\ubbr(t,\boldomega) = \left[ \begin{array}{c}
\bbr(t,\boldomega)\\
\bbr^{\ast}(t,\boldomega)
\end{array} \right], \quad 
\bbr(t,\boldomega) = \left[ 
\def\arraystretch{0.9}
\begin{array}{c}
\bbr(t,\omega_{1})\\
\vdots\\
\bbr(t,\omega_{M})
\end{array} \right] \label{eq:time-spectrum_representation}
\end{equation}
Similarly, the \textit{augmented spectral mean}, $\ubbm(\boldomega) \in \domC^{2MN}$, defined as
\begin{equation}
\ubbm(\boldomega) =  \expect{\ubbr(t,\boldomega)} = \left[\begin{array}{c}
\bbm(\boldomega)\\
\bbm^{\ast}(\boldomega)
\end{array}\right], \quad \bbm(\boldomega) = \left[
\def\arraystretch{0.9}
\begin{array}{c}
\bbm(\omega_{1})\\
\vdots\\
\bbm(\omega_{N})
\end{array}\right]
\end{equation}
while the \textit{augmented spectral covariance}, $\ubbR(\boldomega)\in\domC^{2MN \times 2MN}$, is given by
\begin{equation}
	\begin{aligned}[b]
		\ubbR(\boldomega) &= \cov{\ubbr(t,\boldomega)} = \left[\begin{array}{cc}
		\bbR(\boldomega) & \bbP(\boldomega)\\
		\bbP^{\ast}(\boldomega) & \bbR^{\ast}(\boldomega)
		\end{array}\right] %\label{eq:augmented_spectral_covariance} 
		\\
		\bbR(\boldomega) & = \left[
		\def\arraystretch{0.9}
		\begin{array}{ccc}
		\bbR(\omega_{1}) & \cdots & \bbR(\omega_{1},\omega_{M}) \\
		\vdots & \ddots & \vdots \\
		\bbR(\omega_{M},\omega_{1}) & \cdots & \bbR(\omega_{M}) 
		\end{array}\right] %\label{eq:spectrum_covariance} 
		\\
		\bbP(\boldomega) & = \left[
		\def\arraystretch{0.9}
		\begin{array}{ccc}
		\bbP(\omega_{1}) & \cdots & \bbP(\omega_{1},\omega_{M}) \\
		\vdots & \ddots & \vdots \\
		\bbP(\omega_{M},\omega_{1}) & \cdots & \bbP(\omega_{M})
		\end{array}\right] %\label{eq:spectrum_pseudo-covariance}
	\end{aligned}
\end{equation}

Finally, we arrive at the least-squares estimates of the augmented spectral moments \cite{Scalzo2021}, in the form
\begin{align}
\hat{\ubbm}(\boldomega) & = \frac{1}{T} \sum_{t=0}^{T-1} \uboldPhi^{\Her}(t, \boldomega) \r(t) \label{eq:ML_approx_mean} \\
\hat{\ubbR}(\boldomega) & = \frac{1}{T} \sum_{t=0}^{T-1} \uboldPhi^{\Her}(t, \boldomega) \hat{\s}(t)\hat{\s}^{\Trans}(t) \uboldPhi(t, \boldomega) \label{eq:ML_approx_cov}
\end{align}
with $\hat{\s}(t) = \r(t) - \hat{\m}(t) = \r(t) - \uboldPhi(t, \boldomega)\hat{\ubbm}(\boldomega)$.

\subsection{Graph-Theoretic Diversification}

\subsubsection{Graph Signal Processing}

Following the notation in \cite{12}, we define a graph  $\mathcal{G}=\{\mathcal{V},\mathcal{B} \}$ as being composed of a set of vertices $\mathcal{V}$, which are connected through a set of edges, $\mathcal{B}=\mathcal{V}\times\mathcal{V}$, where the symbol $\times$ denotes a direct product operator. 

The connectivity of a graph, $\mathcal{G}$, is described through a \textit{weight matrix}, $\mathbf{W}\in\mathbb{R}^{N\times N}$, the elements of which are non-negative real numbers, which designate the connection strength between the vertices $m$ and $n$, so that

\begin{equation}
W_{mn}
\begin{cases}
> 0 &  \text{if } (m,n)\in\mathcal{B} \\
= 0 &  \text{if } (m,n)\notin\mathcal{B}
\end{cases}
\end{equation}
The \textit{degree matrix}, $\mathbf{D}\in\mathbb{R}^{N\times N}$, is a diagonal matrix whose diagonal elements, $D_{nn}$, are equal to the sum of weights of all edges connected to a vertex $n$ in an undirected graph
\begin{equation}
D_{nn} = \sum_{m=1}^{N}W_{nm}
\end{equation}
while the \textit{Laplacian matrix} is given by
\begin{equation}
\mathbf{L} = \mathbf{D} - \mathbf{W}
\end{equation}
\subsubsection{Market Graph}
A universe of $N$ assets can be modeled as a \textit{market graph}, with the weight matrix defined as
\begin{equation}
\mathbf{W} 
= 
\begin{bmatrix}
1 & \frac{|\sigma_{12}|}{\sqrt{\sigma_{11}\sigma_{22}}}  & \hdots & \frac{|\sigma_{1N}|}{\sqrt{\sigma_{11}\sigma_{NN}}}  \\
\frac{|\sigma_{21}|}{\sqrt{\sigma_{11}\sigma_{22}}}  & 1 & \hdots &\frac{|\sigma_{2N}|}{\sqrt{\sigma_{22}\sigma_{NN}}}  \\
\vdots & \vdots & \ddots & \vdots \\
\frac{|\sigma_{N1}|}{\sqrt{\sigma_{NN}\sigma_{11}}}  & \frac{|\sigma_{N2}|}{\sqrt{\sigma_{NN}\sigma_{22}}}  & \hdots & 1  \\
\end{bmatrix}
\label{eq:23}
\end{equation}
where $\sigma_{nm}$ denotes the covariance between the returns of asset $n$ and asset $m$. Note the symmetry of the weight matrix, that is, $\sigma_{nm}$ = $\sigma_{mn}$.
%\begin{remark}
%	Note that we have that $\sigma_{nm}$ = $\sigma_{mn}$, which preserves the symmetry of the weight matrix.
%\end{remark}
\begin{figure}[ht]
	\centering
	\includegraphics[height =3.5cm, width=0.5\linewidth]{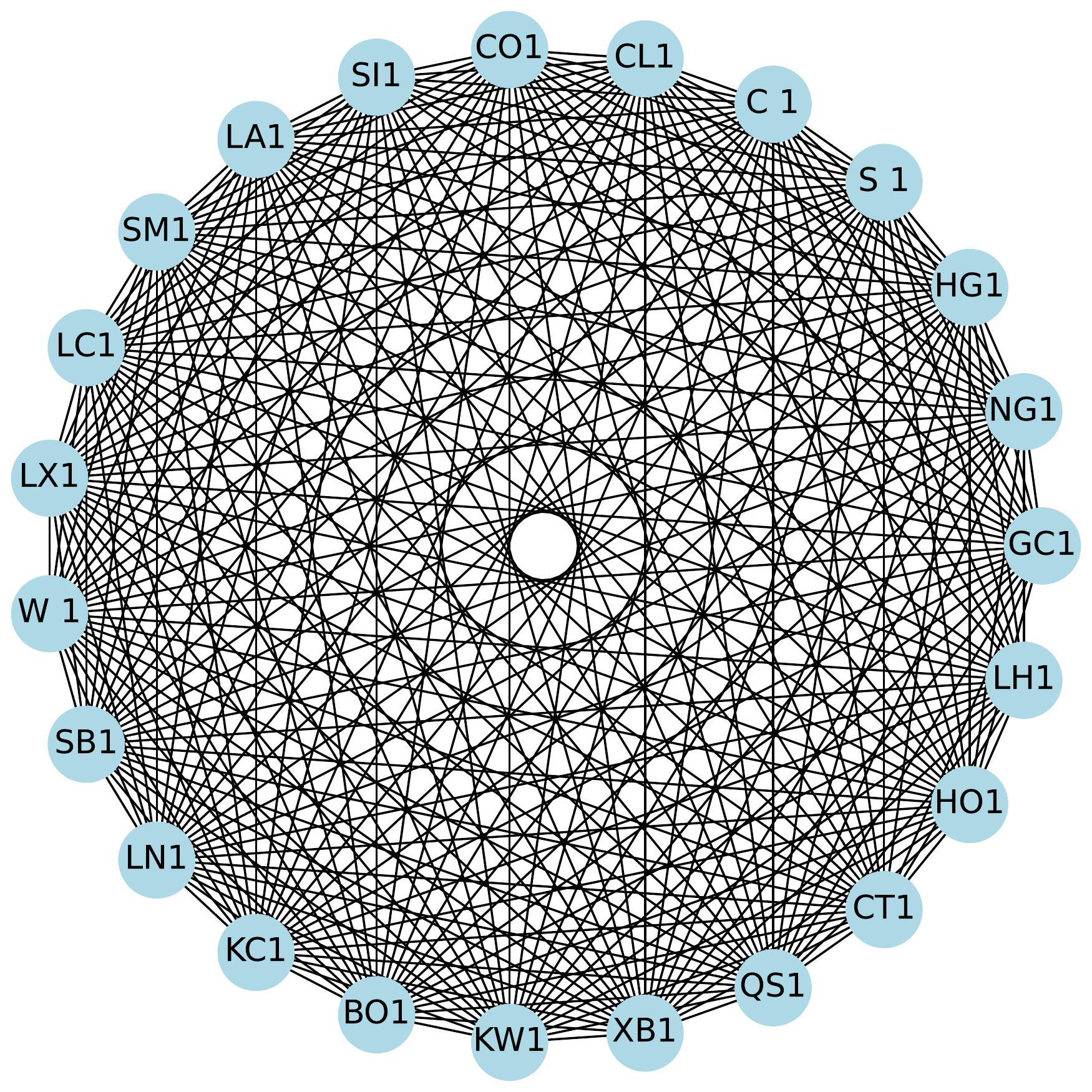}
	\caption{Market graph formed from assets in the \textit{Bloomberg Commodity Index}.}
	\label{fig:1}
\end{figure}

Fig. \ref{fig:1} illustrates one of the fundamental problems when using the covariance matrix in the context of financial investment, as it assumes full vertex connectivity, and thus does not appropriately account for real-world market structure  \cite{LopezdePrado2016}\cite{22}\cite{LopezdePrado2014}\cite{LopezdePrado2014_2}.

\subsubsection{Vertex Clustering and Minimum Cuts}

In order to allow for the clustering of asset vertices into distinct subgroups, we shall introduce vertex clustering based on \textit{minimum cuts}.

Given an undirected graph, $\mathcal{G}$, defined by set of vertices, $\mathcal{V}$, and edge weights, $\mathcal{W}$, we desire to group the vertices of the graph into two subsets, $\mathcal{E}$ and $\mathcal{H}$, such that $\mathcal{E}\subset\mathcal{V}$, $\mathcal{H}\subset\mathcal{V}$, $\mathcal{E}\cup\mathcal{H}=\mathcal{V}$ and  $\mathcal{E}\cap\mathcal{H}=\emptyset$. To this end, a cut of graph, $\mathcal{G}$, given the subset of vertices $\mathcal{E}$ and $\mathcal{H}$ is given by \cite{12}:

\begin{equation}
Cut(\mathcal{E},\mathcal{H}) 
=
\sum_{m\in\mathcal{E}, n\in\mathcal{H}} W_{mn}
\end{equation}
A \textit{minimum cut} is then the cut with the minimal sum of weights joining subsets $\mathcal{E}$ and $\mathcal{B}$. Note that finding the minimal cut in a graph is a combinatorial problem, and thus computationally prohibitive for large graph topologies.

%\begin{remark}
%	Note that finding the minimal cut in a graph is a combinatorial problem in the worst case, and thus can become computationally prohibitive for large graph topologies.
%\end{remark}

In the context of asset allocation in portfolios, it is often desirable that sub-graphs are as a large as possible, to prevent large disparity in asset splits. This motivates the definition of a \textit{normalised ratio cut}, which takes the form \cite{Hagen1992}

\begin{equation}
CutN(\mathcal{E},\mathcal{H}) 
=
\bigg(\frac{1}{N_{\mathcal{E}}} + \frac{1}{N_{\mathcal{H}}}\bigg)
\sum_{m\in\mathcal{E}, n\in\mathcal{H}} W_{mn}
\label{eq:27}
\end{equation}
where $N_\mathcal{E}$ and $N_\mathcal{H}$ represent is the number of elements in subsets $\mathcal{E}$ and $\mathcal{H}$, respectively. The first step to obtaining a computationally tractable way of performing minimum-cut-based vertex clustering is through the notion of of an \textit{indicator vector}, $\mathbf{x}\in\mathbb{R}^{N}$. The elements of an indicator vector are \textit{sub-graph-wise constant}, with the constant values within each cluster of vertices, but distinct across clusters. This implies that $\mathbf{x}$ may serve to uniquely identify the assumed cut of the graph into disjoint subsets \cite{22}, as \textit{e.g.} in the case of two sub-graphs \cite{12}

\begin{equation}
x(n) 
=
\begin{cases}
\frac{1}{N_\mathcal{E}}, & \text{if } n\in\mathcal{E} \\
-\frac{1}{N_\mathcal{H}}, & \text{if } n\in\mathcal{H}
\end{cases} 
\end{equation}
The normalized cut defined in \eqref{eq:27}, can be written in terms of the graph Laplacian and indicator vector as
\begin{equation}
CutN(\mathcal{E},\mathcal{H}) = \frac{\mathbf{x}^{\text{T}}\mathbf{L}\mathbf{x}}{\mathbf{x}^{\text{T}}\mathbf{x}}
\end{equation}
so that the normalized cut can be considered as a minimization problem 
\begin{equation}
\begin{aligned}[b]
\min_{\mathbf{x}} &  \hspace{2mm} \mathbf{x}^{\text{T}}\mathbf{L}\mathbf{x} \\
\text{s.t.} & \hspace{2mm} \mathbf{x}^{\text{T}}\mathbf{x} = 1
\end{aligned}
\end{equation}

The solution to the above problem is given by $\mathbf{x}_{\text{opt}}=\mathbf{u}_{1}$, \cite{12} the second eigenvector of the graph Laplacian, $\mathbf{L}$, also known as the \textit{Fiedler} eigenvector \cite{Fiedler1973}. 

\section{Dynamic Spectral Portfolio Cuts}

Based on the above graph-theoretic interpretation of financial markets, we proceed to introduce a \textit{dynamic market graph}, based on the time-varying covariance matrix presented in \eqref{eq:10}. To this end, we first define the \textit{dynamic weight matrix} as
\begin{equation}
\begin{aligned}[b]
&\mathbf{W}(t) = 
\mathbf{V}(t)\big|\mathbf{R}(t) \big|\mathbf{V}^{\text{T}}(t)
\end{aligned}
\label{eq:29}
\end{equation}
where $\mathbf{V}(t)$ is a diagonal matrix containing the inverse square root of the diagonal elements in $\mathbf{R}(t)$ at a given time instant, $t$, in accordance with the definition in \eqref{eq:23}. Note that the modulus operator $|\cdot|$ is applied element-wise. This time-varying generalisation of the market graph makes it possible to capture economic cycles and shocks, thus allowing for a more meaningful and informative analysis of asset relationships.

In the context of graph data analytics, time-varying graph matrices naturally give rise to the concept of \textit{dynamic graph matrix spectra}, whereby the eigenspectrum and eigenspace also become nonstationary, and have embedded information on the cyclical relationships captured by $\mathbf{R}(t)$. Mathematically, the singular value decomposition of the graph weight matrix in \eqref{eq:29} now becomes

\begin{equation}
	\mathbf{W}(t) = \mathbf{U}(t)\mathbf{\Lambda}(t)\mathbf{U}^{\text{T}}(t)
\end{equation}
given that $\mathbf{W}(t)$ it is a symmetric square invertible matrix. The eigenvector and eigenvalue matrices, $\mathbf{U}(t)$ and $\mathbf{\Lambda}(t)$, are in turn respectively given by

\begin{equation}
\mathbf{U}(t) 
=
\begin{bmatrix}
\mathbf{u}_1(t)  & \mathbf{u}_2(t) & \hdots & \mathbf{u}_N(t) 
\end{bmatrix}
\end{equation}

\begin{equation}
\mathbf{\Lambda}(t) 
=
\begin{bmatrix}
\lambda_1(t) & 0 & \hdots & 0 \\
0  & \lambda_2(t) & \hdots & 0 \\
\vdots  & \vdots & \ddots & \vdots \\
0  & 0 & \hdots & \lambda_N(t)  \\
\end{bmatrix}
\end{equation}

This allows us to carry out a time-varying extention of the operations traditionally performed on graphs, which we refer to as \textit{dynamic graph data analytics}, which includes the notions of time-varying clustering or vertex dimensionality reduction. In the context of the market graph, this would imply clustering assets into different sub-graphs at each time instant, $t$, thereby modelling more accurately the seasonal economical relationships between assets across a business year.

Following the capital allocation scheme proposed in \cite{22}, we denote by $h_i$ the percentage of capital allocated to $\mathcal{G}_i$, and consider two cases:

\begin{enumerate}
	\item $h_i = \frac{1}{2^{K_i}}$, where $K_i$ represents the number of cuts made to the market graph to obtain the cluster in question;
	\item $h_i = \frac{1}{K+1}$, where $K$ represents the number of individual clusters generated through the cuts.
\end{enumerate}
\vspace{-3mm}
\begin{figure}[h!]
	\centering
	\begin{minipage}[b]{0.49\linewidth}
		\centering
		\includegraphics[height = 2cm, width=0.48\textwidth]{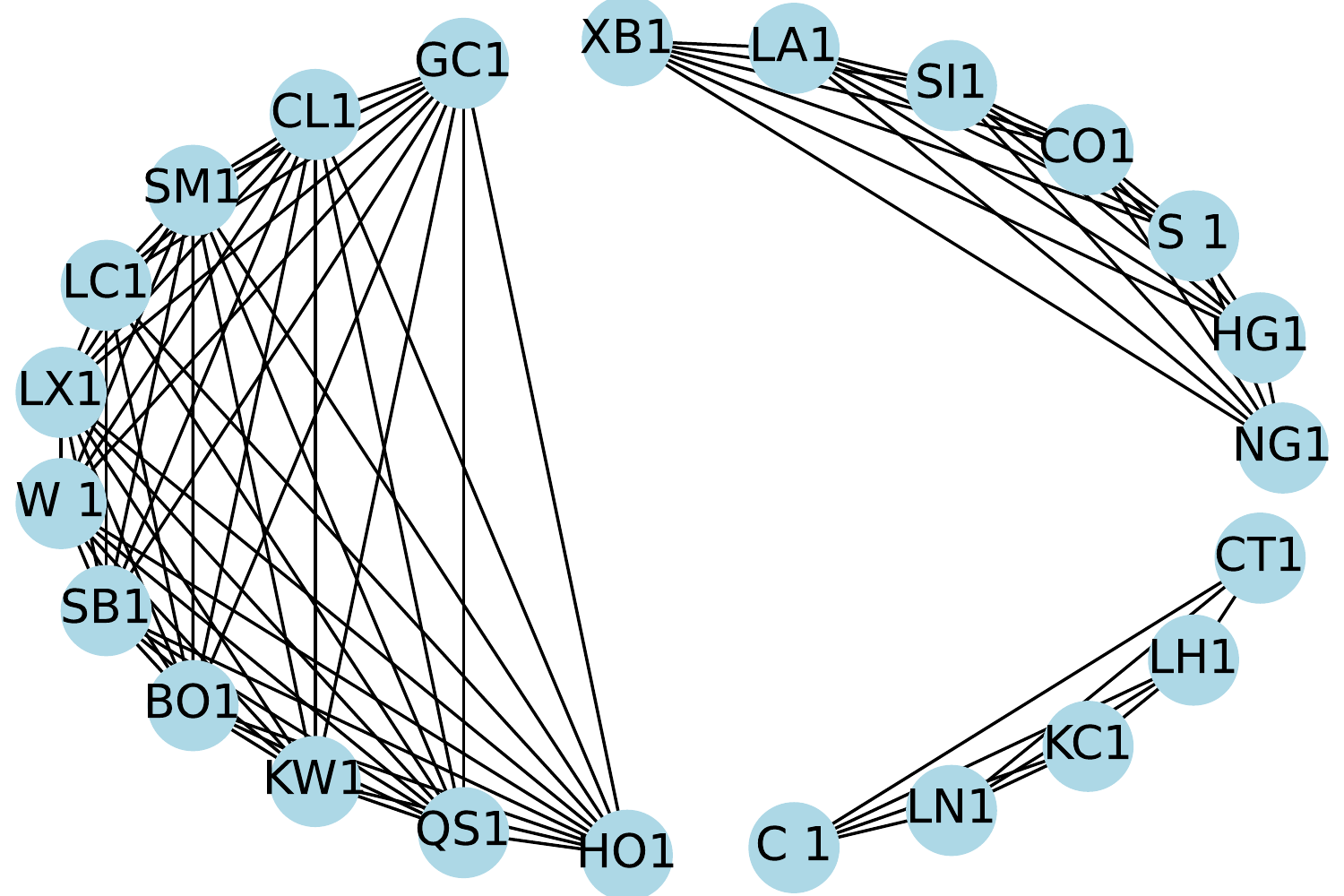}
		\vspace{4mm}
		%\caption{$t=1$, $K=2$ cut.\vspace{4mm}}
	\end{minipage}
	\hfill
	\begin{minipage}[b]{0.49\linewidth}
		\centering
		\includegraphics[height = 2cm, width=0.48\textwidth]{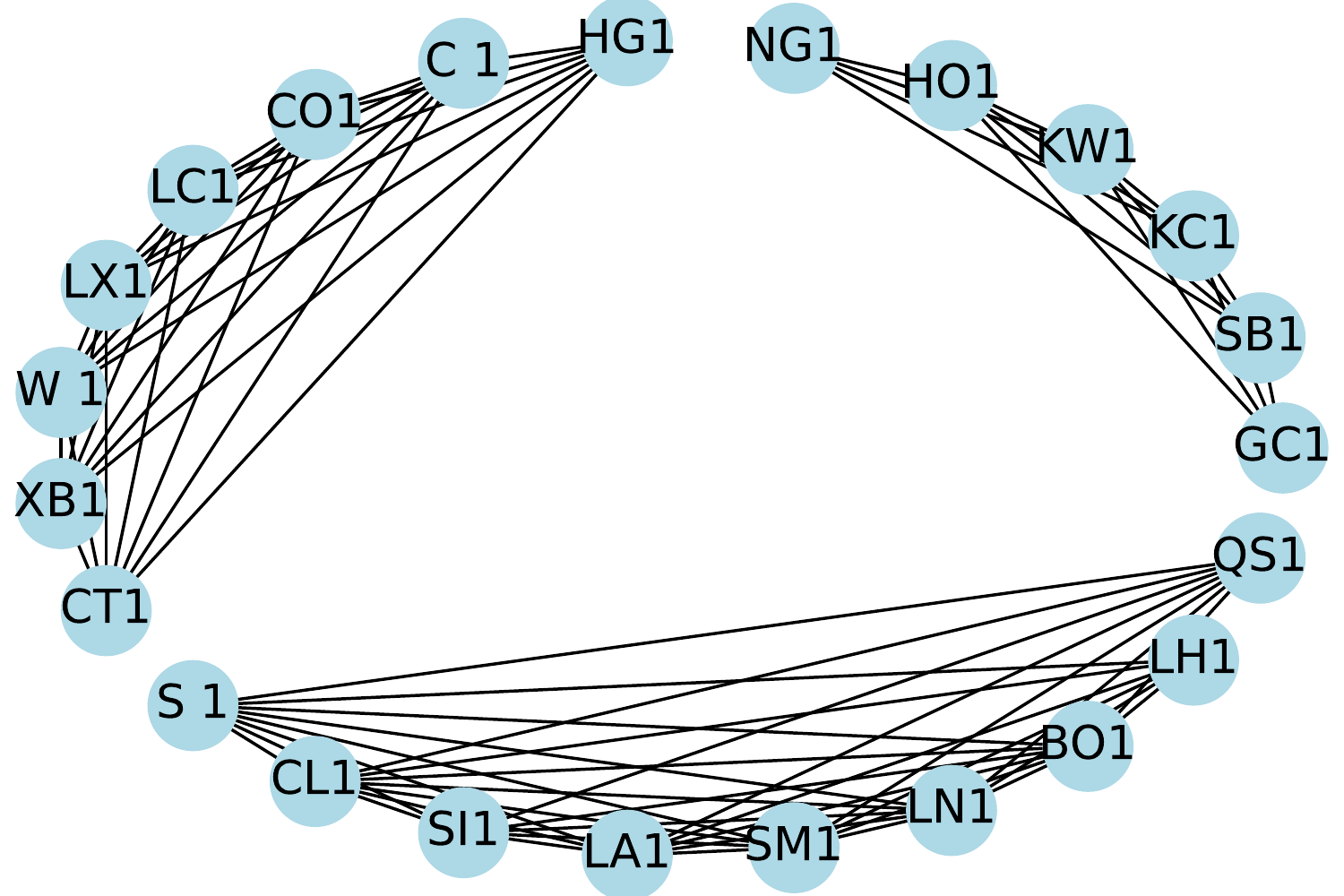}
		\vspace{4mm}
		%\caption{$t=2$, $K=2$ cut.\vspace{4mm}}
	\end{minipage}
	\begin{minipage}[b]{0.49\linewidth}
		\centering
		\includegraphics[height = 2cm, width=0.48\textwidth]{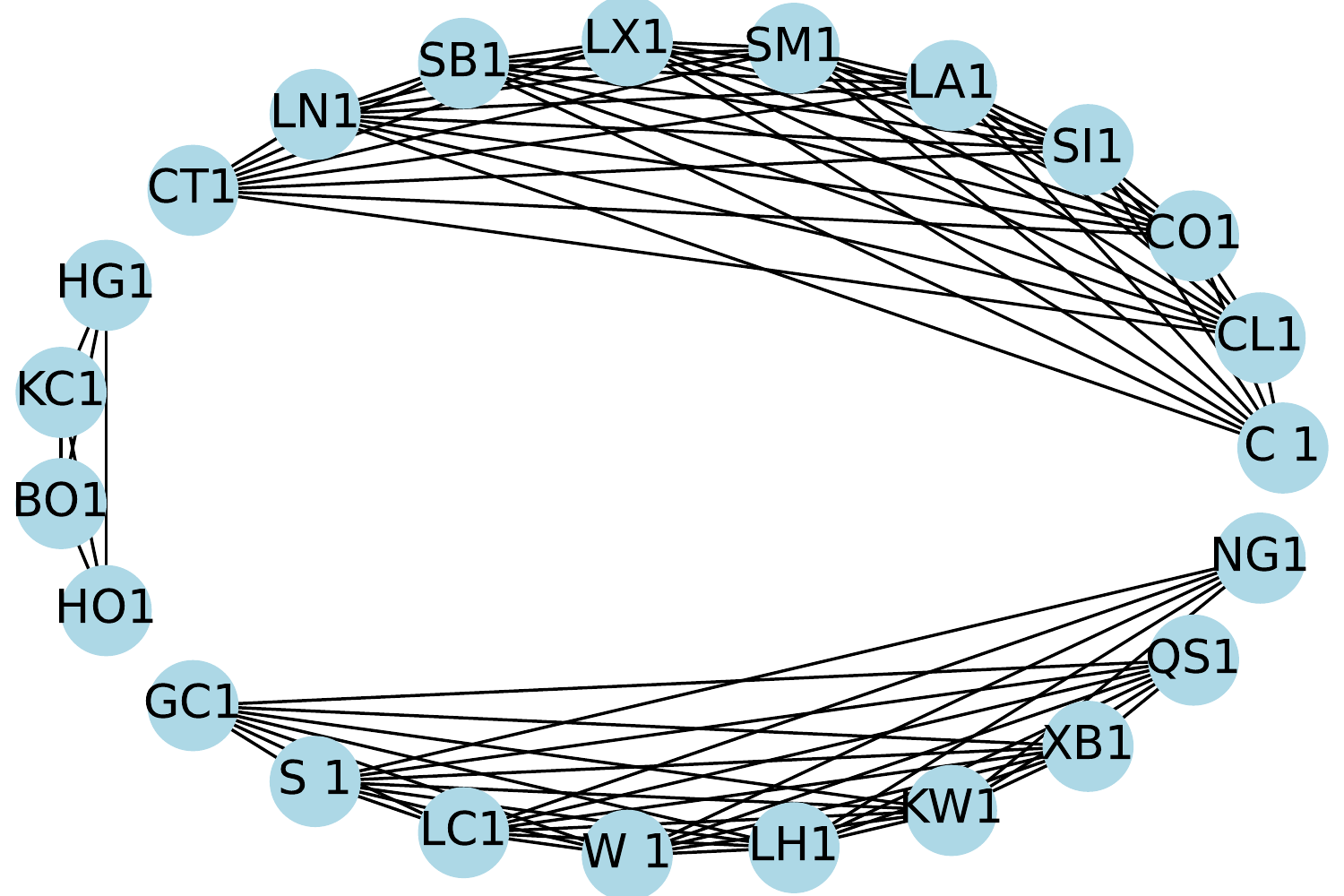}		
		\vspace{2mm}
		%\caption{$t=3$, $K=2$ cut.}
	\end{minipage}
	\hfill
	\begin{minipage}[b]{0.49\linewidth}
		\centering
		\includegraphics[height = 2cm, width=0.48\textwidth]{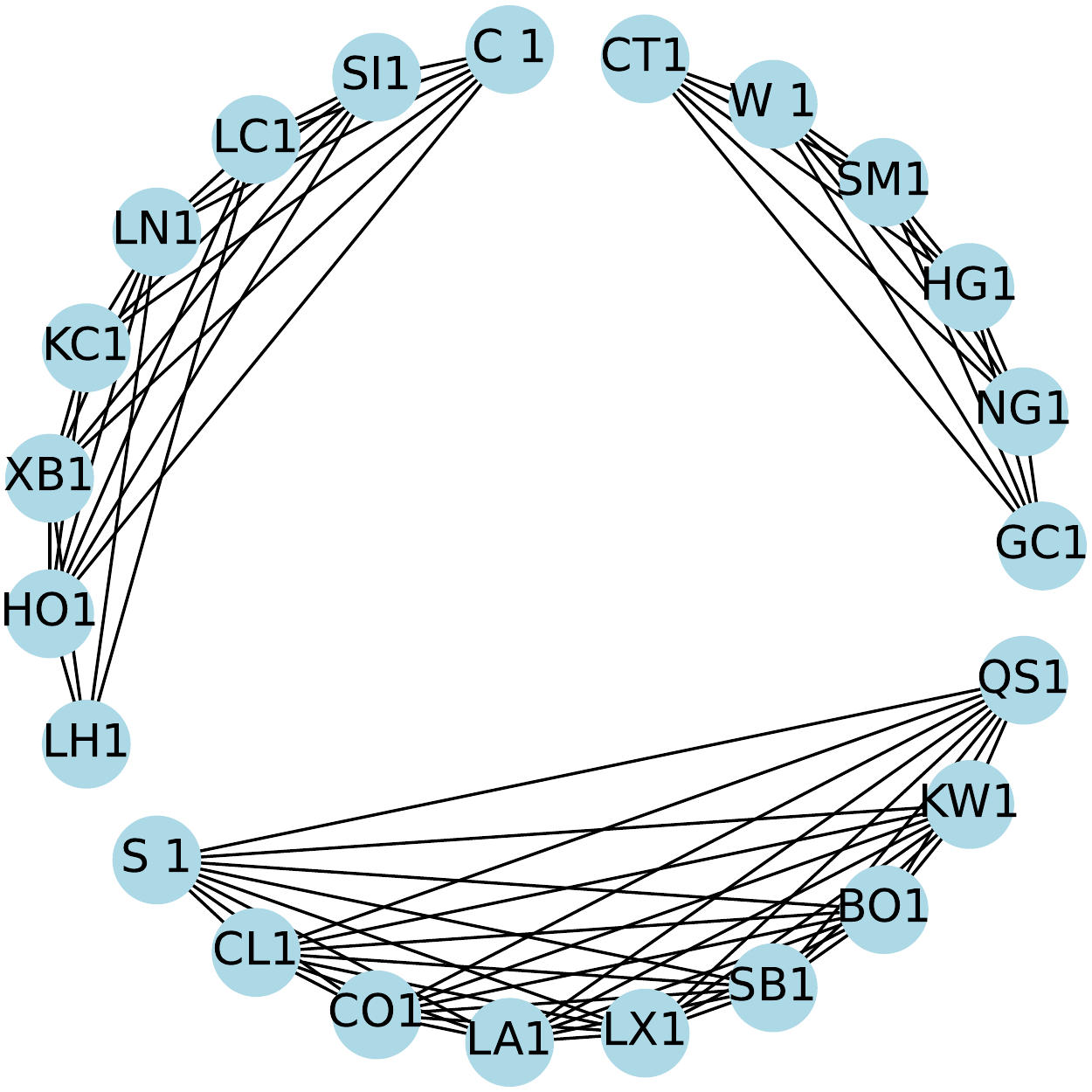}
		\vspace{2mm}
		%\caption{$t=4$, $K=2$ cut.}
	\end{minipage}
	\caption{Example of $K=2$ minimum cuts performed on the the \textit{Bloomberg Commodity Index }(BCOM) market graph, with 23 vertices and at 4 different time instants. Note that the dynamic nature of the graph weight matrix results in different cluster formations being generated at each time instant, $t$.}
	\label{fig:2}
\end{figure}

%\begin{remark}
%It should be mentioned that the number of cuts, $K$, is a \textit{hyperparameter} of choice. Given that the number of asset clusters is not known in advance, this quantity needs to be chosen based on the performance of the model in a validation set.
%\end{remark}
\vspace{-5mm}
\section{Simulations}
\begin{figure}[h!]
	\centering
	\begin{minipage}[b]{0.5\textwidth}
		\centering
		\includegraphics[height = 2.75cm, width=0.85\textwidth]{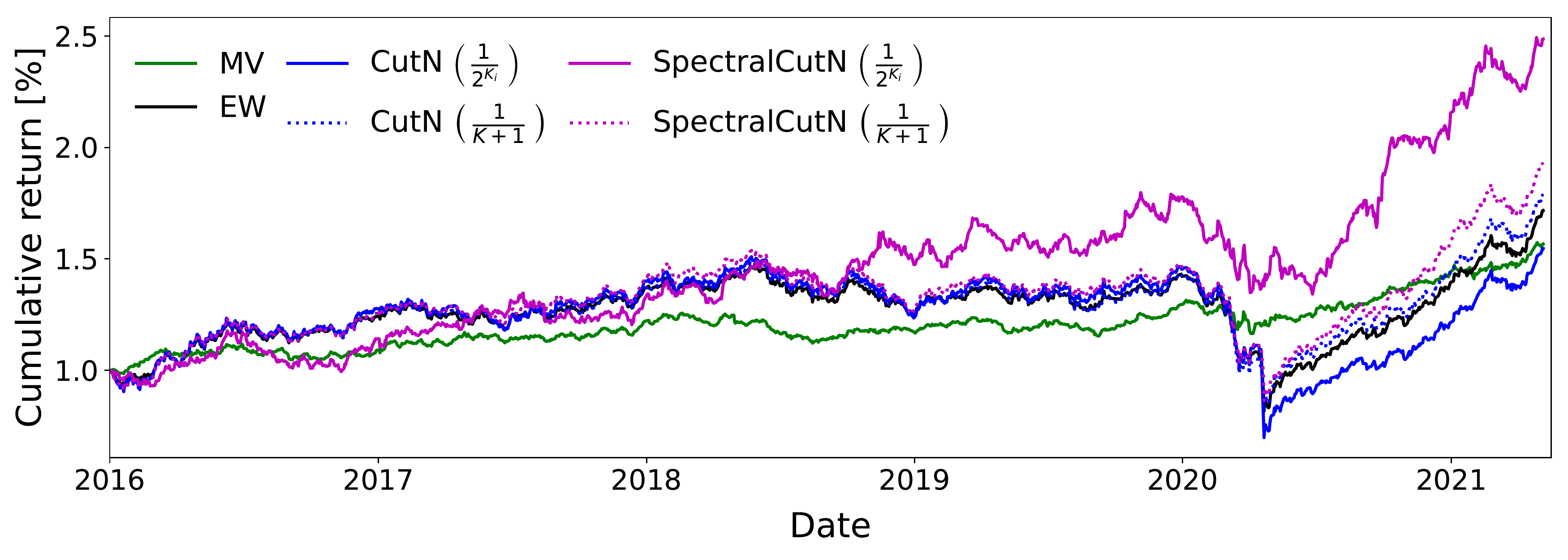}
		%\vspace{-1mm}
		%\caption{BCOM Index, $K=15$ graph cuts.}
	\end{minipage}
	\hfill
	\begin{minipage}[b]{0.5\textwidth}
		\centering
		\includegraphics[height = 2.75cm, width=0.85\textwidth]{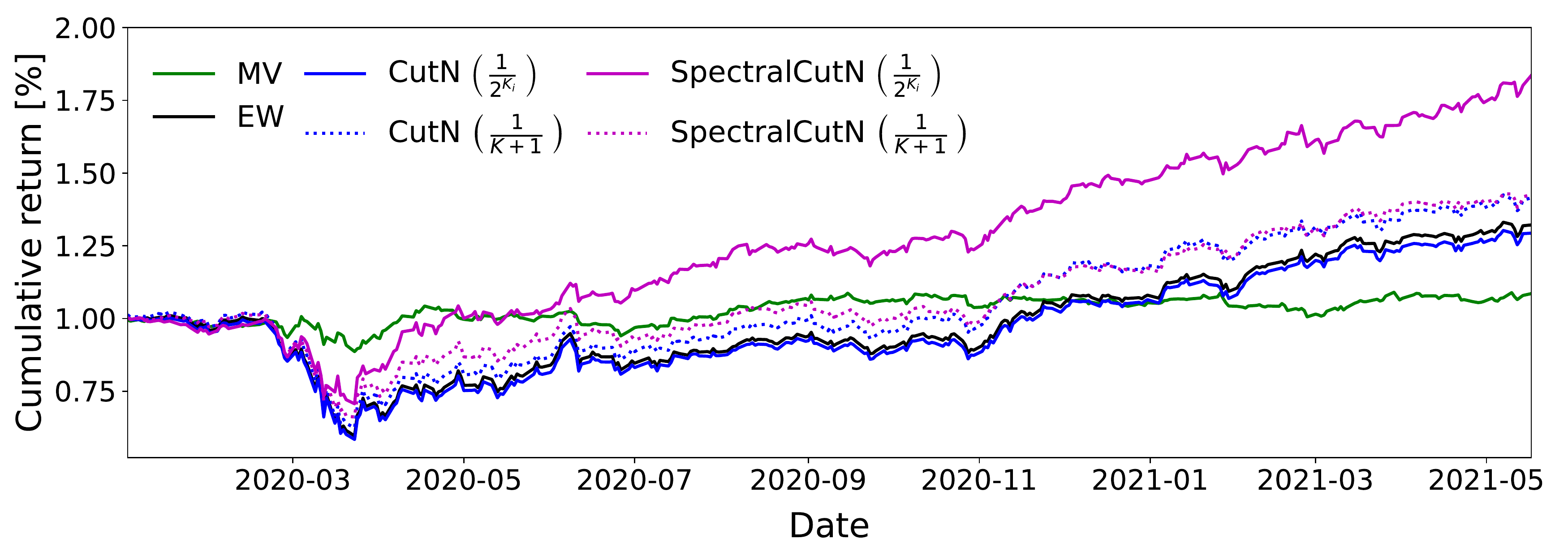}
		%\vspace{-1mm}
		%\caption{S\&P 500 Index, $K=10$ graph cuts.}
	\end{minipage}
	%\vspace{-7mm}
	\caption{Out-of-sample performance of all strategies on BCOM index (top) and S\&P 500 (bottom) for $K=15$ and $K=10$ market graph cuts respectively.}
	\label{fig:3}
\end{figure}

%\vspace{-1mm}
The performance of the proposed dynamic portfolio cuts framework was investigated using historical price data of the $23$ commodity futures contracts constituting the Bloomberg Commodity Index in the period 2010-01-01 to 2021-05-17, as well as the 100 most liquid stocks in the S\&P 500
index, based on average trading volume, between 2015-01-01
to 2021-05-17. The data was partitioned into a \textit{training} (in-sample) dataset, with dates 2010-01-01 to 2016-01-01 for the BCOM index and 2014-01-02 to 2020-01-02 for the S\&P 500, which was used to estimate the spectral covariance and retrieve its time-varying counterpart. Subsequently, asset clustering was carried out on the dynamic market graph and tested on data from the \textit{test} (out-of-sample) dataset, with dates 2016-01-01 to 2021-05-17 for the BCOM index and 2020-01-02 to 2021-05-17 for the S\&P 500. Fig. \ref{fig:3} shows a comparison between the proposed dynamic portfolio cut and its static counterpart, as well as standard equally-weighted (EW) and MVO portfolios.  

\vspace{10cm}

The results shown in Fig. \ref{fig:3} show that the proposed dynamic spectral cuts framework consistently results in a larger cumulative return compared to both standard and existing graph-based approaches. As desired, the so enabled high average returns, coupled with the low variance of the proposed strategy, result in higher \textit{Sharpe ratios}, as summarized in Tables \ref{table:1} and \ref{table:2}.

\begin{table}[h!]
	\caption{Sharpe ratio over a varying number of cuts $K$ in BCOM Index.}
	\vspace{-3mm}
	\scriptsize
	\renewcommand{\arraystretch}{1}
	\setlength{\tabcolsep}{1pt}
	\centering % used for centering table
	\begin{tabular}{c c c c c c c c c } % centered columns (3 columns)
		\hline\hline %inserts double horizontal lines
		\textbf{Strategy} & \textbf{Allocation}  & $K=1$ &  $K=2$ & $K=3$ & $K=4$ & $K=5$ & $K=10$ & $K=15$ \\ [0.5ex] % inserts table
		%heading
		\hline % inserts single horizontal line 
		SpectralCutN & $\frac{1}{2^{K_i}}$ & \textbf{2.15} & 2.77 & \textbf{2.7} & \textbf{2.73} & \textbf{2.72} & \textbf{2.89} & \textbf{3.19} \\ % inserting body of the table
		SpectralCutN & $\frac{1}{K+1}$ & \textbf{2.15} & \textbf{2.88} & 2.51  & 2.37  & 1.75 & 2.22 & 1.71 \\
		CutN & $\frac{1}{2^{K_i}}$ & 1.96 & 1.14 & 1.12 & 2.22 & 2.08 & 0.75 & 1.07  \\
		CutN & $\frac{1}{K+1}$ & 1.62 & 1.81 & 1.86 & 1.85 & 1.99 & 1.78 & 1.3 \\ [0.5ex] % [1ex] adds vertical space
		\hline %inserts single line
	\end{tabular}
	\label{table:1} % is used to refer this table in the text
\end{table}
\begin{table}[h!]
	\caption{Sharpe ratio over a varying number of cuts $K$ in S\&P 500.}
	\vspace{-3mm}
	\scriptsize
	\renewcommand{\arraystretch}{1}
	\setlength{\tabcolsep}{1pt}
	\centering % used for centering table
	\begin{tabular}{c c c c c c c c c } % centered columns (3 columns)
		\hline\hline %inserts double horizontal lines
		\textbf{Strategy} & \textbf{Allocation}  & $K=1$ &  $K=2$ & $K=3$ & $K=4$ & $K=5$ & $K=10$ & $K=50$ \\ [0.5ex] % inserts table
		%heading
		\hline % inserts single horizontal line 
		SpectralCutN & $\frac{1}{2^{K_i}}$ & 1.61 & \textbf{1.78} & \textbf{1.86} & \textbf{1.87}  & \textbf{1.87} & \textbf{1.88} & \textbf{1.76} \\ % inserting body of the table
		SpectralCutN & $\frac{1}{K+1}$ & 1.61 & 1.6 & 1.51  & 1.37  & 1.21 & 1.12 & 0.97 \\
		CutN & $\frac{1}{2^{K_i}}$ & 0.86  & 0.81 & 0.94 & 0.86 & 0.84 & 0.82 & 0.85 \\
		CutN & $\frac{1}{K+1}$ & \textbf{1.63} & 1.5 & 1.23 & 1.35 & 1.25 & 1.05 & 0.91 \\ [0.5ex] % [1ex] adds vertical space
		\hline %inserts single line
	\end{tabular}
	\label{table:2} % is used to refer this table in the text
\end{table}

Note that graph-based portfolio strategies inevitably result in long-only portfolios, given the positive weights connecting the vertices of a graph. As such, portfolio cuts and dynamic portfolio cuts are expected to work well on upward trending indices, such as the S\&P 500, which are only composed of stocks, and have a tendency to grow over time.

\section{Conclusions}

A novel dynamic spectral graph framework has been introduced which allows to model the interaction of financial assets residing on the market graph over time. This is achieved through a class of spectral estimators of the augmented spectral covariance, which is shown to account for cyclostationary trends in market data, and thus economic cycles and shocks. Simulations have demonstrated the advantages of the proposed framework over stationary portfolio cut techniques on the market graph, as well as a dominant performance over traditional portfolio optimization approaches.

\bibliography{Bibliography.bib}
\bibliographystyle{IEEEtran}

\end{document}